\documentclass[aps,prl,twocolumn,superscriptaddress]{revtex4}
\usepackage{amsmath}
\usepackage{amssymb}
\usepackage{graphicx}

\usepackage{color}
\usepackage{bm}
\usepackage{epstopdf}

\begin{document}

\title{Density-matrix approach for an interacting polariton system}

\author{I. G. Savenko}
\affiliation{Science Institute, University of Iceland, Dunhagi 3,
IS-107, Reykjavik, Iceland}
\affiliation{Academic University - Nanotechnology Research and Education Centre, 8/3 Khlopina, 195220, St.Petersburg, Russia}

\author{E.B. Magnusson}
\affiliation{Science Institute, University of Iceland, Dunhagi 3,
IS-107, Reykjavik, Iceland}

\author{I. A. Shelykh}
\affiliation{Science Institute, University of Iceland, Dunhagi 3,
IS-107, Reykjavik, Iceland}
\affiliation{Academic University - Nanotechnology Research and Education Centre, 8/3 Khlopina, 195220, St.Petersburg, Russia}
\date{\today}

\begin{abstract}
Using Lindblad approach we develop a general formalism for
theoretical description of a spatially inhomogenous bosonic system
with dissipation provided by the interaction of bosons with a phonon
bath. We apply our results to model the dynamics of an
interacting one dimensional polariton system in real space and time,
analyzing in detail the role of polariton-polariton and polariton-
phonon interactions.
\end{abstract}

\maketitle

\section{I. Introduction}

A semiconductor microcavity is a photonic
structure designed to enhance the light-matter interaction. In a
planar microcavity photons are confined between two mirrors and
resonantly interact with the excitonic transition of a two-dimensional
semiconductor quantum well (QW). If the quality factor of the cavity
is sufficiently high, it is possible to achieve the regime of strong
coupling between the cavity photon and QW exciton. In this case the
elementary excitations in the system, which are called cavity
polaritons, have a hybrid, half-light, half-matter nature. The
peculiar properties of polaritons make them a unique laboratory
for studying of various collective phenomena interesting from the
point of view of basic physics, which range from polariton BEC
\cite{KasprzakNature}, superfluidity \cite{AmoNature} and Josephson
effect \cite{LagoudakisJosephson} to polariton-mediated
superconductivity \cite{LaussySupercond}.

Besides fundamental interest, quantum microcavities in the strong
coupling regime can be used for a variety of optoelectronic
applications \cite{Imamoglu}. Recently, it was proposed that the
peculiar spin structure and possibility to achieve lateral
confinement of polaritons \cite{Confinement} opens a way for
creation of optical analogs of spintronic components (so-called
spinoptronic devices), based on transport of cavity polaritons in
real space. In this context, the analysis of one-dimensional
(1D) polariton  transport is of particular importance
\cite{WertzNature}, as 1D polariton channels are fundamental
building blocks of such future spinoptronic devices as polariton
neurons \cite{LiewNeuron} and polariton integrated circuits
\cite{LiewCircuit}.

On the theoretical side, transport properties of exciton-polaritons
in real space have not yet been studied in detail. Early works
\cite{SpinHall,Glazov,ShelykhBerry} treated the case of polaritons
interacting with external potentials only neglecting both
polariton-polariton and polariton-phonon interactions. However, in
the most interesting regime of polariton BEC neither of the two
interactions can in principle be neglected. Indeed, coupling of the
polaritons with a reservoir of acoustic phonons leads to
thermalization of the polariton subsystem, which is dramatically
speeded up by polariton- polariton interactions which are known to
be responsible for overcoming the so-called "bottleneck effect"
\cite{Bottleneck}. Besides, in the regime of polariton BEC
polariton-polariton interactions are responsible for the onset of
superfluidity \cite{AmoNature}.

Currently, there are two ways of describing a system of interacting
polaritons. Assuming full coherence of the polariton
system, polariton-polariton interactions can be accounted for using
nonlinear Gross- Pitaevskii (GP) equation, satisfactorily describing
the dynamics of inhomogeneous polariton droplets in real space and
time \cite{Carusotto,ShelykhGP}. The approach, however, does not
include interactions with a phonon bath, responsible for
thermalization of the system and leading to dephasing. In the
opposite limit, when the polariton system is supposed to be fully
incoherent, its dynamics can be described using a system of
semiclassical Boltzmann equations
\cite{Porras2002,Kasprzak2008,Haug2005,Cao}, which provides information about
time dependence of the occupation numbers in reciprocal space
but fails to describe real space dynamics in the inhomogeneous
system.

Recently, there appeared theoretical attempts to combine the two
mentioned approaches introducing dissipation terms into GP equation
in a phenomenological way \cite{Wouters2007}. To our mind, however,
these attempts, although being interesting and leading to a rich
phenomenology \cite{BerloffVortex,Tim1D}, are not fully
satisfactory, as they lack any microscopic justification. In the
present paper we give an alternative way of describing the dynamics
of an inhomogeneous polariton system in real space and time
accounting for dissipation effects. Our consideration is based on
Lindblad approach for density matrix dynamics. We use our results
for modeling of the propagation of polariton droplets in 1D
channels. It should be noted, that the method we develop is rather
general and can in principle be applied to any system of interacting
bosons in contact with a phonon reservoir (for example, indirect
excitons \cite{Butov}). We have previously applied this formalism to model Josephson oscillations between two spatially separated condensates, but in that context the spatial dependence was trivial \cite{Josephson2010}.

\section{II. Formalism}

We describe the state of the system (polaritons
plus phonons) by its density matrix $\rho$, for which we apply
Born approximation factorizing it into the phonon part which is supposed
to be time-independent and corresponds to the thermal distribution
of acoustic phonons
$\rho_{ph}=\texttt{exp}\left\{-\beta\widehat{H}_{ph}\right\}$ and
the polariton part $\rho_{pol}$ whose time dependence should be
determined,  $\rho=\rho_{ph}\otimes \rho_{pol}$.  Our aim is to find
 dynamic equations for the time evolution of the single-particle
polariton density matrix in real space.
\begin{eqnarray}
\rho(\textbf{r},\textbf{r}',t)=Tr\left\{\widehat{\psi}^\dagger(\textbf{r},t)\widehat{\psi}(\textbf{r}',t)\rho\right\}
\equiv\langle\widehat{\psi}^\dagger(\textbf{r},t)\widehat{\psi}(\textbf{r}',t)\rangle
\end{eqnarray}
where
$\widehat{\psi}^\dagger(\textbf{r},t),\widehat{\psi}(\textbf{r},t)$
are operators of the polariton field, and the trace is performed by
all degrees of freedom of the system. The particularly interesting
quantities are the diagonal matrix elements which give the density
of the polariton field in real space and time
$n(\textbf{r},t)=\rho(\textbf{r},\textbf{r},t)$. In our
consideration we neglect the spin of the cavity polaritons for
simplicity, as our main goal here is to account for dissipative
dynamics coming from interaction with phonons, and spin degree of
freedom is not expected to introduce any qualitatively new physics
from this point of view. However, that
introduction of spin into the model is straightforward and
corresponding work is currently underway. It should also be noted that our model is to some extent similar to one proposed in Ref.~\onlinecite{WoutersWigner}. Differently from that paper, however, we assume that all decoherence in the system comes from the interaction with acoustic phonons which is neglected in Ref.~\onlinecite{WoutersWigner} and thus do not perform the separation of the polariton ensemble into coherent low energy part and incoherent high- energy reservoir.

It is convenient to go to reciprocal space, making a Fourier transform of the one-particle density matrix,

\begin{align}
\rho(\textbf{k},\textbf{k}',t)&=(2\pi)^d/L^d\int e^{i(\textbf{kr}-\textbf{k}'\textbf{r}')}\rho(\textbf{r},\textbf{r}',t)d\textbf{r}d\textbf{r}'=\\
\nonumber&=Tr\left\{a_{\textbf{k}}^+a_\textbf{k+q}\right\}\equiv\langle a_{\textbf{k}}^+a_{\textbf{k}'}\rangle
\end{align}
where $d$ is the dimensionality of the system ($d=2$ for non-confined
polaritons, $d=1$ for the polariton channel), L is its
linear size,  $a_{\textbf{k}}^+$, $a_{\textbf{k}}$ are creation and
annihilation operators of the polaritons with momentum \textbf{k}.
Note, that we have chosen the prefactor in a Fourier transform in
such a way, that the values of $\rho(\textbf{k},\textbf{k}',t)$ are
dimensionless, and diagonal matrix elements give occupation numbers
of the states in discretized reciprocal space. Knowing the density
matrix in reciprocal space, we can find the density matrix in real
space straightforwardly applying inverse Fourier transform.

The total Hamiltonian of the system can be represented as a sum of two parts,
\begin{equation}
H=H_1+H_2
\end{equation}
where the first term $H_1$ describes the "coherent" part of the
evolution, corresponding to free polariton propagation and
polariton-polariton interactions and the second term $H_{phon}$
corresponds to the dissipative interaction with acoustic phonons.
The two terms affect the polariton density matrix in a qualitatively
different way.

\subsection{A. Polariton-polariton interactions} 

The part of the evolution
corresponding to $H_1$ is given by the following expression
\begin{equation}
H_1=\sum_{\textbf{k}} E_\textbf{k}  a_{\textbf{k}}^+a_\textbf{k}+
\frac{U}{2}\sum_{\textbf{k}_1,\textbf{k}_2,\textbf{p}}a_{\textbf{k}_1}^+a_{\textbf{k}_2}^+a_{\textbf{k}_1+\textbf{p}}a_{\textbf{k}_2-\textbf{p}}
\end{equation}
where $E_\textbf k$ is the energy dispersion of the polaritons, $U$
is the matrix element of the polariton-polariton interactions. 
In the current paper we neglect the \textbf{p}-dependence of the polariton- polariton interaction constant coming from Hopfield coefficients \cite{polpol}. We do this approximation because the goal of the manuscript is to present a novel formalism for description of the relaxation effects and not detailed modeling of a particular experiment, and we want to keep our formalism as simple as possible. Besides, this approximation is widely used in current description of polaritonic systems based on modifications of the Gross-Pitaevskii equations \cite{Wouters2007,BerloffVortex,Tim1D}. However, for modeling of realistic experiments the \textbf{p}-dependence of the interaction constant can easily be introduced into the equations.
 
The effect of $H_1$ on the evolution of the density matrix is described by the
Liouville-von Neumann equation,
\begin{equation}
i\left(\partial_t\rho\right)^{(1)}=\left[H_1;\rho\right]
\label{liouville}
\end{equation}
which after use of mean field approximation leads to the
following dynamic equations for the elements of the single-particle
density matrix in reciprocal space (see Appendix I for details of the derivation):
\begin{align}
&\left\{\partial_t\rho(\textbf{k},\textbf{k})\right\}^{(1)}=
 2U\sum_{\textbf{k}_1,\textbf{p}}\textrm{Im}\left\{\rho(\textbf{k}_1,\textbf{k}_1-\textbf{p})\rho(\textbf{k},\textbf{k}+\textbf{p})\right\}
\label{EqOccupancyPolPol}\\
&\left\{\partial_t\rho(\textbf{k},\textbf{k}')\right\}^{(1)}
=i(E_\textbf{k}-E_\textbf{k}')\rho(\textbf{k},\textbf{k}')+ \label{EqCoherencePolPol} \\
\nonumber &+iU\sum_{\textbf{k}_1,\textbf{p}}\rho(\textbf{k}_1,\textbf{k}_1-\textbf{p})\left[\rho(\textbf{k}-\textbf{p},\textbf{k}')-\rho(\textbf{k},\textbf{k}'+\textbf{p})\right]
\end{align}

These expressions represent an analog of the Gross-Pitaevskii
equation written for the density matrix.

\subsection{B. Scattering with acoustic phonons}

Polariton-phonon scattering corresponds to the interaction of the
quantum polariton system with a classical phonon reservoir. It is of
dissipative nature, and thus straightforward application of the
Liouville-von Neumann equation is impossible. Introduction of
dissipation into quantum systems is an old problem, for which there
is no single well established solution. In the domain of quantum
optics, however, there are standard methods based on the Master
Equation techniques \cite{Carmichael}. In the following we give a
brief outline of this approach applied to a dissipative polariton
system.

The Hamiltonian of interaction of polaritons with acoustic
phonons in Dirac representation can be represented as
\begin{align}
&\widehat{H}_2(t)=H^-(t)+H^+(t)=\\
\nonumber&=\sum_{\textbf{k},\textbf{q}}D(\textbf{q})e^{i(E_{\textbf{k}+\textbf{q}}-E_{\textbf{k}})t}a_{\textbf{k}+\textbf{q}}^+a_\textbf{k}(b_\textbf{q}e^{-i\omega_{\textbf{q}}t}+b_{-\textbf{q}}^+e^{i\omega_{\textbf{q}}t})
\end{align}
where $a_\textbf{k}$ are operators for polaritons, $b_\textbf{q}$
operators for phonons, $E_\textbf{k}$ and $\omega_\textbf{q}$ are
dispersion relations for polaritons and acoustic phonons
respectively, $D(\textbf{q})$ is the polariton-phonon coupling
constant. In the last equality we separated the terms $H^+$ where a
phonon is created, containing the operators $b^+$, from the terms
$H^-$ in which it is destroyed, containing operators $b$.

Now, one can consider a hypothetical situation when
polariton-polariton interactions are absent, and all redistribution
of the polaritons in reciprocal space is due to the scattering with
a thermal reservoir of acoustic phonons. One can rewrite the
Liouville-von Neumann equation in an integro-differential form and
apply the so called Markovian approximation, corresponding to the
situation of fast phase memory loss (see Ref.~\onlinecite{Carmichael}
for the details and discussion of limits of validity of the
approximation)
\begin{align}
\left(\partial_t\rho\right)^{(2)}&=-\int_{-\infty}^t dt'\left[H_{2}(t);
  \left[H_{2}(t');\rho(t)\right]\right]= \label{Liouville_int}\\
\nonumber &= \delta_{\Delta E}\left[2\left(H^+\rho H^-+H^-\rho\right.
H^+\right)-   \nonumber \\
&\left.-\left(H^+H^-+H^-H^+\right)\rho-\rho\left(H^+H^-+H^-H^+\right)\right] \nonumber
\end{align}
where the coefficient $\delta_{\Delta E}$ denotes energy conservation and has dimensionality of inverse energy and in the calculation taken to be equal to the broadening of the polariton state \cite{KavokinMalpuech}.
For time evolution of the mean value of any arbitrary operator $ \langle \widehat{A}\rangle=Tr(\rho
\widehat{A})$ due to scattering with phonons one thus has (derivation of this formula is represented in Appendix II):
\begin{equation}\label{eqM}
\left\{\partial_t\langle
\widehat{A}\rangle\right\}^{(2)}=\delta_{\Delta E}\left(\langle[H^-;[\widehat{A};H^+]]\rangle+\langle[H^+;[\widehat{A};H^-]]\rangle\right).
\end{equation}
Putting $\widehat{A}=a_\textbf{k}^+a_{\textbf{k}'}$  in this equation
we get the contributions to the dynamic equations for the elements
of the single-particle density matrix coming from polariton-phonon
interaction (see Appendix III):
\begin{align}
 \label{BoltzmannOcc}
\left\{\partial_t\rho(\textbf{k},\textbf{k})\right\}^{(2)}&=\\
\nonumber
\sum_{\textbf{q}',E_\textbf{k}<E_{\textbf{k+q}'}}&2W(\textbf{q}')\left[\rho(\textbf{k+q}',\textbf{k+q}')(n_{\textbf{q}'}^{ph}+1)(\rho(\textbf{k},\textbf{k})+1) \right.\\
\nonumber
&\left.-\rho(\textbf{k},\textbf{k})n_{\textbf{q}'}^{ph}(\rho(\textbf{k+q}',\textbf{k+q}')+1)\right]+ \\
\nonumber
+\sum_{\textbf{q}',E_\textbf{k}>E_{\textbf{k+q}'}}&2W(\textbf{q}')\left[\rho(\textbf{k+q}',\textbf{k+q}')n_{-\textbf{q}'}^{ph}(\rho(\textbf{k},\textbf{k})+1)-\right. \\
\nonumber
&\left.-\rho(\textbf{k},\textbf{k})(n_{-\textbf{q}'}^{ph}+1)(\rho(\textbf{k+q}',\textbf{k+q}')+1)\right]
\end{align}
and
\begin{align}
\label{BoltzmannCorr}
&\left\{\partial_t\rho(\textbf{k},\textbf{k}')\right\}^{(2)}=\rho(\textbf{k},\textbf{k}') \times \\\nonumber
&\times \left\{\sum_{\textbf{q}',E_\textbf{k}<E_{\textbf{k+q}'}}W(\textbf{q}')\left[\rho(\textbf{k+q}',\textbf{k+q}')-n_{\textbf{q}'}^{ph}\right]+ \right. \\\nonumber
+&\sum_{\textbf{q}',E_\textbf{k}>E_{\textbf{k+q}'}}W(\textbf{q}')\left[-\rho(\textbf{k+q}',\textbf{k+q}')-n_{\textbf{q}'}^{ph}-1\right]+\\
\nonumber
+&\sum_{\textbf{q}',E_\textbf{k}'<E_{\textbf{k}'+\textbf{q}'}}W(\textbf{q}')\left[\rho(\textbf{k}'+\textbf{q}',\textbf{k}'+\textbf{q}')-n_{\textbf{q}'}^{ph}\right] +\\\nonumber
+&\left.\sum_{\textbf{q}',E_\textbf{k}'>E_{\textbf{k}'+\textbf{q}'}}W(\textbf{q}')\left[-\rho(\textbf{k}'+\textbf{q}',\textbf{k}'+\textbf{q}')-n_{\textbf{q}'}^{ph}-1\right]\right\}
\end{align}
where the transition rates are given by
$W(\textbf{q})=D^2(\textbf{q})\delta_{\Delta E_\textbf{q}}$ and
$n_\textbf{q}^{ph}$ denote the occupation numbers of phonons
with wavevector $\textbf{q}$ given by Bose distribution.

Eq.~\eqref{BoltzmannOcc} is nothing but a standard Boltzmann
equation for the phonon-assisted polariton relaxation, while
Eq.~\eqref{BoltzmannCorr} describes the decay of the off-diagonal
matrix elements of a single-particle density matrix due to
interaction with classical phonon reservoir. Together these equations
thus describe thermalization of the polariton system.

To account for the effects of free polariton propagation, polariton-polariton
and polariton-phonon interactions one should combine
together expressions \eqref{EqOccupancyPolPol},
\eqref{EqCoherencePolPol}, \eqref{BoltzmannOcc} and
\eqref{BoltzmannCorr}. After finding the single-particle density
matrix in reciprocal space by solving the corresponding dynamic
equations, one can determine the dynamic of the system in real space
simply performing a Fourier transformation by \textbf{k} and
\textbf{k}' variables.

\section{III. Results and discussion} 

Our formalism is suitable for the
description of both 2D polaritons and polaritons confined within 1D
channels. In this paper we present the results of numerical modeling
for the latter case only, as solving of the dynamic equations for 2D
case requires the use of supercomputing facilities.

\begin{figure}
\includegraphics[width=0.85\linewidth]{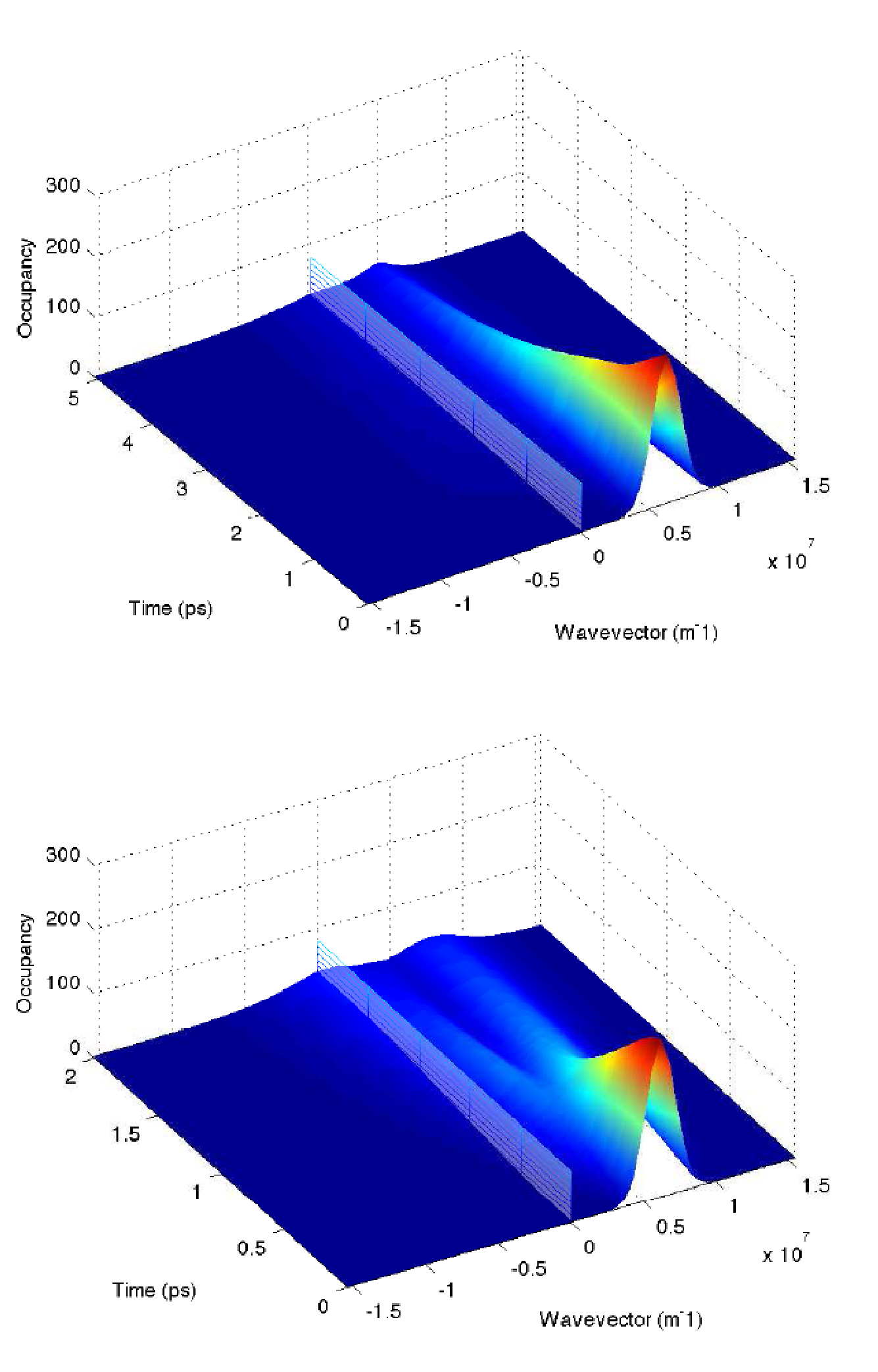}
\caption{Time evolution of polariton distribution in $k$-space. The
vertical plane denotes $k=0$ state. a) Only polariton-phonon
scattering is accounted for. The relaxation towards $k=0$ state is
blocked due to the bottleneck effect. b) Both polariton-phonon and
polariton-polariton scattering are accounted for. The maximum of the
polariton concentration is developed at $k=0$, signifying the
overcoming of the bottleneck effect. Another maximum appears at
higher $k$, due to the energy-conserving nature of the polariton-polariton
interactions. } \label{fig:1}
\end{figure}

The results of modeling are shown on Figures 1-3. We consider a $2$
$\mu$m wide polariton channel in GaAs microcavity with Rabi
splitting 15 meV at temperature $T=1$~K. The polaritons are created
by a short coherent localized laser pulse. We account for the finite
lifetime of cavity polaritons $\tau=5$ ps adding the term
$-\rho(\textbf{k},\textbf{k}')/\tau$ into the dynamic equations.

Figure 1 shows the dynamics of the polariton system in reciprocal
space and demonstrates the roles of the polariton-phonon and
polariton-polariton interactions. If only the former are included,
the system demonstrates the bottleneck effect shown on Fig.~1a. Due
to the energy relaxation coming from the interactions with phonons,
polaritons have a tendency to move towards the ground state in
$k$-space. However, this process is dramatically slowed down in the
inflection region of the polariton dispersion, where
polariton-phonon interaction becomes inefficient. Consequently,
there is no remarkable increase of the population of $k=0$ state
\cite{Bottleneck}. The bottleneck effect can be overcome by the
polariton-polariton interactions, as shown on Fig.~1b. One sees that
in this case the particles accumulate quickly in $k=0$ state. At the
same time, the second maximum of the polariton distibution appears
at higher $k$ due to the energy conservative nature of the
polariton-polariton interactions (analogically to the formation of
the idler mode in polariton parametric oscillator \cite{PPO}).

\begin{figure}
\includegraphics[width=0.9\linewidth]{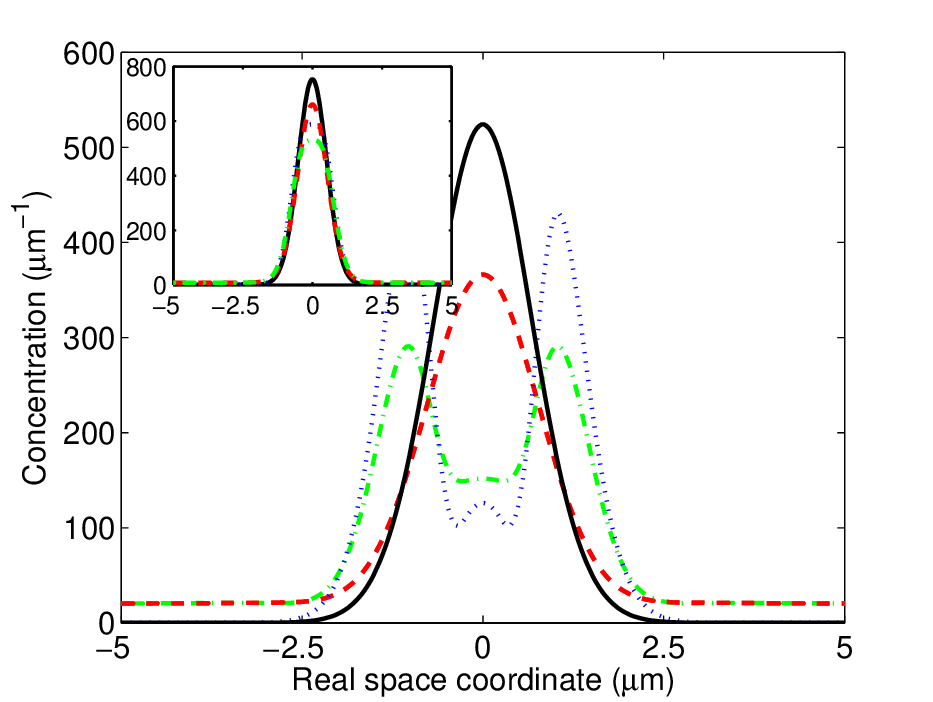}
\caption{Polariton distribution in real space at times $t=0.5$ ps
(inset) and $t=1.5$ ps after creating the package by
a localized laser pulse centered around $k=0$. Black solid line
corresponds to ballistic propagation, red dashed line - to polariton-phonon
interactions, blue dotted line - to polariton-polariton
interactions and green dashed line - to both polariton-phonon and
polariton-polariton interactions.}
\end{figure}

Our results for the dynamics of the polariton distribution in
reciprocal space are in good agreement with those obtained by using
Boltzmann equations. In addition, our approach allows consideration
of the dynamics of the dissipative polariton system in real space.
This is illustrated on Figures 2 and 3.

Figure 2 shows the effect of the various types of interactions on
the real space dynamics of the localized polariton wavepackage. We
compare the cases of the ballistic propagation with those where only
polariton-phonon interactions are included, only polariton-polariton
interactions are included and both polariton-polariton and
polariton-phonon interactions are included. As one sees, the
dynamics are very different for these four cases.
Polariton-polariton interactions lead to splitting of the
wavepackage into two soliton- like peaks, which is in good
qualitative agreement with the results given by Gross-Pitaevskii
equation \cite{ShelykhGP}. On the other hand, polariton-phonon
interactions lead to damping of the package, contributing to the
recovering of the homogenious distribution of the polaritons in real
space as it is expected from the classical diffusion equation.

\begin{figure}
\includegraphics[width=0.9\linewidth]{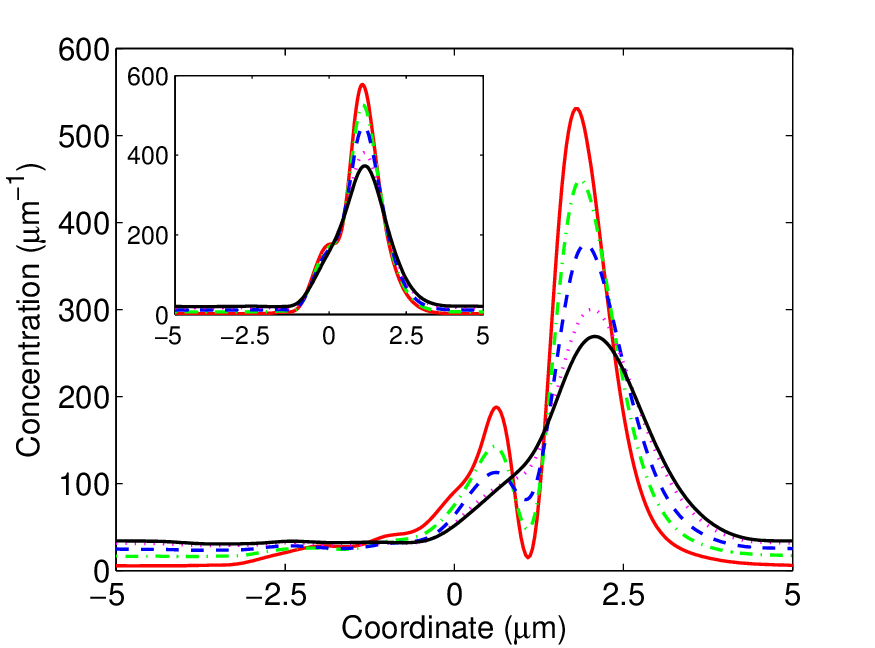}
\caption{Polariton distribution in real space at times $t=0.5$ ps (inset)
and $t=1.5$ ps after creating the package by a
localized laser pulse centered around $k=3\cdot10^6$ m$^{-1}$.
Scattering on phonons and polaritons are both accounted for. The
temperatures are: 1 K (red/solid), 4 K (green/dash-dot), 8 K
(blue/dashed), 15 K (magenta/dotted) and 20 K (black/solid).}
\end{figure}

Naturally, the effect of the phonon damping strongly depends on
temperature, as shown on Fig.~3. One sees, that at low
temperatures the propagating package is split into two due to the
polariton-polariton interactions. Increasing the temperature
smoothes down the polariton distibution and at $T=20K$ one has just
a single peak.

\section{IV. Conclusion}

In conclusion, we developed a formalism for the description of the
dissipative dynamics of an inhomogeneous polariton system in real
space and time accounting for polariton-polariton interactions and
polariton-phonon scattering. The formalism was applied for numerical
modeling of the propagation of a polariton droplet in a 1D channel.
Our approach can also be used for modeling the dissipative dynamics
of other bosonic systems (e.g. indirect excitons).

\section{Acknowledgements}

We thank G. Malpuech, D.D. Solnyshkov and T.C.H. Liew for useful
discussions. The work was supported by Rannis "Center of excellence
in polaritonics" and FP7 IRSES projects "SPINMET" and "POLAPHEN".


\section{Appendix I: derivation of kinetic equations for
polariton- polariton scattering}

The evaluation of the dynamic equations of $\langle a_\textbf k ^\dagger a_\textbf k\rangle$ and $\langle a_\textbf k ^\dagger a_\textbf {k+q}\rangle$ using Eq.~\eqref{liouville} and applying the mean field approximation
gives:

\begin{widetext}
\begin{align}
\textrm{Tr}\left\{a_\textbf{k}^+a_{\textbf{k+q}}\left[\rho;H_{p-p}\right]\right\}&=
U\sum_{\textbf{k}_1,\textbf{k}_2,\textbf{p}}\rho(\textbf{k}_2,\textbf{k}_2-\textbf{p})\textrm{Tr}\left\{a_\textbf{k}^+a_{\textbf{k+q}}\left[\rho;a_{\textbf{k}_1}^+a_{\textbf{k}_1+\textbf{p}}\right]\right\}=\\
&=U\sum_{\textbf{k}_2,\textbf{p}}\rho(\textbf{k}_2,\textbf{k}_2-\textbf{p})\left[-\rho(\textbf{k},\textbf{k+q+p})+\rho(\textbf{k-p},\textbf{k+q})\right] \nonumber
\end{align}

\begin{align}
\textrm{Tr}\left\{a_\textbf{k}^+a_{\textbf{k}}\left[\rho;H_{p-p}\right]\right\}=
U\sum_{\textbf{k}_1,\textbf{k}_2,\textbf{p}}\rho(\textbf{k}_2,\textbf{k}_2-\textbf{p})\textrm{Tr}\left\{a_\textbf{k}^+a_{\textbf{k}}\left[\rho;a_{\textbf{k}_1}^+a_{\textbf{k}_1+\textbf{p}}\right]\right\}=\\
\nonumber
=U\sum_{\textbf{k}_1,\textbf{k}_2,\textbf{p}}\rho(\textbf{k}_2,\textbf{k}_2-\textbf{p})\textrm{Tr}\left\{\rho
\left[a_{\textbf{k}_1}^+a_{\textbf{k}_1+\textbf{p}}a_\textbf{k}^+a_{\textbf{k}}-
a_\textbf{k}^+a_{\textbf{k}}a_{\textbf{k}_1}^+a_{\textbf{k}_1+\textbf{p}}\right]\right\}=-2iU\sum_{\textbf{k}_2,\textbf{p}}\textrm{Im}\left\{\rho(\textbf{k}_2,\textbf{k}_2-\textbf{p})\rho(\textbf{k},\textbf{k+p})\right\} \nonumber
\end{align}

From this one straightforwardly gets
Eqs.~\ref{EqOccupancyPolPol},~\ref{EqCoherencePolPol}.
\end{widetext}

\section{Appendix II: Derivation of expression for dynamics of mean values in Born- Markov approximation}

Now, consider the evolution of a mean value of any arbitrarty
operator $A, \langle A\rangle=Tr(\rho A)$ (energy conserving
delta-function omitted) if dynamics of a density matrix is given by
Eq.~\ref{Liouville_int}:

\begin{widetext}
\begin{align}
\partial_t\langle A\rangle&=\texttt{Tr}\left(2H^+\rho H^-A+2H^-\rho
H^+A-\left(H^+H^-+H^-H^+\right)\rho
A-\rho\left(H^+H^-+H^-H^+\right)A\right)=\\
\nonumber
&=\texttt{Tr}\left[\rho\left(2H^-AH^++2H^+AH^--A\left(H^+H^-+H^-H^+\right)-\left(H^+H^-+H^-H^+\right)A\right)\right]
\end{align}
where we used the property of the invariance of the trace as regards
cyclic permutations of operators. The latter expression can be simplified as

\begin{align}
&2H^-AH^++2H^+AH^--A\left(H^+H^-+H^-H^+\right)-\left(H^+H^-+H^-H^+\right)A=\\
\nonumber={}&2H^-([A;H^+]+H^+A)+2H^+([A;H^-]+H^-A)-A\left(H^+H^-+H^-H^+\right)-\left(H^+H^-+H^-H^+\right)A=\\
\nonumber={}&2H^-[A;H^+]+2H^+[A;H^-]-H^+[A;H^-]-[A;H^+]H^--H^-[A;H^+]-[A;+H^-]H^+=\\
\nonumber={}&[H^-;[A;H^+]]+[H^+;[A;H^-]]
\end{align}
\end{widetext}
where we used the following property of commutators:
$[A;BC]=B[A;C]+[A;B]C$. Thus

\begin{equation}
\partial_t\langle A\rangle=\texttt{Tr}\left(\rho[H^-;[A;H^+]]\right)+\texttt{Tr}\left(\rho[H^+;[A;H^-]]\right)
\end{equation}

For the important case of the Hermitian operator A corresponding to
a physical observable one has

\begin{align}
&[H^-;[A;H^+]]^+=-[[A;H^+]^+;H^+]=\\
\nonumber ={}&[[H^-;A^+];H^+]=[H^+;[A;H^-]]
\end{align}
and

\begin{equation}
\partial_t\langle A\rangle=2\texttt{Re}\{\texttt{Tr}\left(\rho[H^-;[A;H^+]]\right)\}
\end{equation}

This formula can be applied for calculation of occupation numbers.

\section{Appendix III: Derivation of dynamic equations with acoustic
phonons} To get explicit expressions for dynamics of
$\rho(\textbf{k},\textbf{k},t)$ let us consider a simple
case when only states \textbf k, \textbf k+\textbf q are present. We have two cases:

a) $E(\textbf k+\textbf q)>E(\textbf k)$. In this case, leaving energy- conserving terms
only one gets
\begin{align}
H^+&=Da_\textbf k^+a_{\textbf k+\textbf q}b_\textbf q^+\\
H^-&=Da_{\textbf k+\textbf q}^+a_\textbf kb_\textbf q 
\end{align}\\

The application of Eq.\ref{eqM} gives the following results:
\begin{align}
&\delta^{-1}\partial_t\rho(\textbf k,\textbf k)=\delta^{-1}\partial_tn_\textbf k=\texttt{Tr}\left(\rho[H^-;[a_\textbf k^+a_\textbf k;H^+]]\right)+\\
\nonumber&\texttt{Tr}\left(\rho[H^+;[a_\textbf k^+a_\textbf k;H^-]]\right)=2\texttt{Tr}\left(\rho[H^+;[a_\textbf k^+a_\textbf k;H^-]]\right)
\end{align}
After some straightforward algebra one gets

\begin{widetext}
\begin{align}
[a_\textbf k^+a_\textbf k;H^-]=D[a_\textbf k^+a_\textbf k;a_{\textbf k+\textbf q}^+a_\textbf kb_\textbf q]=Da_{\textbf k+\textbf q}^+b_\textbf q[a_\textbf k^+a_\textbf k;a_\textbf k]=-Da_{\textbf k+\textbf q}^+a_\textbf kb_\textbf q,\\
\nonumber
[H^+;[a_\textbf k^+a_\textbf k;H^-]]=-D^2[a_\textbf k^+a_{\textbf k+\textbf q}b^+_\textbf q;a_{\textbf k+\textbf q}^+a_\textbf kb_\textbf q]=\\
\nonumber=D^2\left(a_{\textbf k+\textbf q}^+a_{\textbf k+\textbf q}(a_\textbf k^+a_\textbf k+1)(b_\textbf q^+b_\textbf q+1)-(a_{\textbf k+\textbf q}^+a_{\textbf k+\textbf q}+1)a_\textbf k^+a_\textbf kb^+_\textbf qb_\textbf q\right)
\end{align}
and
\begin{equation}
\partial_t\rho(\textbf k,\textbf k)=2\delta D^2\left[\rho(\textbf k+\textbf q,\textbf k+\textbf q)(n_\textbf q^{ph}+1)(\rho(\textbf k,\textbf k)+1)-\rho(\textbf k,\textbf k)n_\textbf q^{ph}(\rho(\textbf k+\textbf q,\textbf k+\textbf q)+1)\right]
\end{equation}
\end{widetext}
This is nothing but ordinary Boltzmann equation, accounting for transition from state $\textbf{k+q}$ to state $\textbf{k}$ accompanied by the emission of the phonon (spontaneous or stimulated) and transition from state $\textbf{k}$ to state $\textbf{k+q}$ due to the phonon absorption.
\\
b) $E(\textbf k+\textbf q)<E(\textbf k)$. Treating this case in a similar way we find:
\begin{widetext}
\begin{equation}
\partial_t\rho(\textbf k,\textbf k)=2\delta D^2\left[\rho(\textbf k+\textbf q,\textbf k+\textbf q)n_{-\textbf q}^{ph}(\rho(\textbf k,\textbf k)+1)-(\rho(\textbf k+\textbf q,\textbf k+\textbf q)+1)(n_{-\textbf q}^{ph}+1)\rho(\textbf k,\textbf k)\right]
\end{equation}
\end{widetext}
Again we obtained Boltzmann equation, but differently to the previous case the transition from  state $\textbf{k+q}$ to state $\textbf{k}$ goes with absorption of the phonon and from  state $\textbf{k}$ to state $\textbf{k+q}$ with phonon emission. Performing summation over all reciprocal space one gets Eq.~\eqref{BoltzmannOcc}.

Now let us consider the dynamics of the off-diagonal part of the density matrix $\rho(\textbf{k},\textbf{k}',t)$.
\begin{align}
&\delta^{-1}\partial_t\rho(\textbf k,\textbf k')=\label{offdiag_phon}\\
\nonumber={}&\texttt{Tr}\left(\rho[H^+;[a_\textbf k^+a_{\textbf k'};H^-]]\right)+\texttt{Tr}\left(\rho[H^-;[a_\textbf k^+a_{\textbf k'};H^+]]\right)
\end{align}

Here we should consider different orderings of the energies corresponding to the states $\textbf k$, $\textbf k'$ and state $\textbf k+\textbf q$. As an example, let us consider the case when  
 $E_\textbf k<E_{\textbf k+\textbf q},\ E_{\textbf k'}<E_{\textbf k+\textbf q}$. Leaving energy-conserving terms only one gets
\begin{align}
H^+&=Da_\textbf k^+a_{\textbf k+\textbf q}b_\textbf q^++a_{\textbf k'}^+a_{\textbf k+\textbf q}b_{\textbf q'}^+\\
H^-&=Da_\textbf ka_{\textbf k+\textbf q}^+b_\textbf q+a_{\textbf k'}a_{\textbf k+\textbf q}^+b_{\textbf q'}
\end{align}
where $\textbf q'=\textbf k+\textbf q-\textbf k'$. In this case the first term in Eq.~\eqref{offdiag_phon} gives:
\begin{widetext}
\begin{align}
[a_{\textbf k}^+a_{{\textbf k}'};H^-]&=D[a_{\textbf k}^+a_{{\textbf k}'};a_{\textbf k+\textbf q}^+a_\textbf kb_\textbf q+a_{\textbf k'}a_{\textbf k+\textbf q}^+b_{\textbf q'}]=-Db_{\textbf q'}a_{\textbf k'}a_{\textbf k+\textbf q}^+\\\
\nonumber
[H^+;[a_\textbf k^+a_{\textbf k'};H^-]]&=D^2[a_{\textbf k+\textbf q}a_\textbf k^+b^+_\textbf q+a_{\textbf k+\textbf q}a_{\textbf k'}^+b^+_{\textbf q'};-b_\textbf qa_{\textbf k+\textbf q}^+a_{\textbf k'}]=\\
\nonumber
&=D^2\left\{a_\textbf k^+a_{\textbf k'}(a_{\textbf k+\textbf q}^+a_{\textbf k+\textbf q}+1)b_\textbf q^+b_\textbf q-(b_\textbf q^+b_\textbf q+1)a_{\textbf k+\textbf q}^+a_{\textbf k+\textbf q}a_\textbf k^+a_\textbf k\right\}\\
\nonumber
\end{align}
and
\begin{align}
[a_{\textbf k}^+a_{{\textbf k}'};H^+]&=D[a_{\textbf k}^+a_{{\textbf k}'};a_{\textbf k+\textbf q}a_\textbf k^+b_\textbf q^++a_{\textbf k'}^+a_{\textbf k+\textbf q}b_{\textbf q'}^+]=Db_{\textbf q'}^+a_{\textbf k}^+a_{\textbf k+\textbf q}\\
\nonumber
[H^-;[a_\textbf k^+a_{\textbf k'};H^+]]&=D^2[a_{\textbf k+\textbf q}^+a_\textbf kb_\textbf q+a_{\textbf k+\textbf q}^+a_{\textbf k'}b_{\textbf q'};b_{\textbf q'}^+a_{\textbf k}^+a_{\textbf k+\textbf q}]=\\
\nonumber
&=D^2\left\{a_{\textbf k+\textbf q}^+a_{\textbf k+\textbf q}a_\textbf k^+a_{\textbf k'}(b_{\textbf q'}^+b_{\textbf q'}+1) - (a_{\textbf k+\textbf q}^+a_{\textbf k+\textbf q}=1)a_\textbf k^+a_{\textbf k'}b_{\textbf q'}^+b_{\textbf q'}\right\}\\
\nonumber
\end{align}
Finally for the case $E_\textbf k<E_{\textbf k+\textbf q},\ E_{\textbf k'}<E_{\textbf k+\textbf q}$ one obtains equation in the form

\begin{align}
\partial_t\rho(\textbf k,\textbf k')={}&\delta
D^2\left(\rho(\textbf k+\textbf q,\textbf k+\textbf q)(n_\textbf q^{ph}+1)-(\rho(\textbf k+\textbf q,\textbf k+\textbf q)+1)n_\textbf q^{ph}\right)\rho(\textbf k,\textbf k')+\\
      & \left(\rho(\textbf k+\textbf q,\textbf k+\textbf q)(n_{\textbf q'}^{ph}+1)-(\rho(\textbf k+\textbf q,\textbf k+\textbf q)+1)n_{\textbf q'}^{ph}\right)\rho(\textbf k,\textbf k')
\end{align}
\end{widetext}

The same procedure is applied for all other cases. Performing again summation over all reciprocal space one gets
Eq.~\ref{BoltzmannCorr}.



\begin{thebibliography}{99}

\bibitem{KasprzakNature} J. Kasprzak, M. Richard, S. Kundermann, A. Baas, P.
Jeambrun, J. M. J. Keeling, F. M. Marchetti, M. H. Szy- manska, R. Andre, J. L. Staehli, V. Savona, P. B. Lit- tlewood, B. Deveaud and Le Si Dang, Nature \textbf{443}, 409 (2006).

\bibitem{AmoNature} A. Amo, D. Sanvitto, F. P. Laussy, D. Ballarini, E. del Valle, M. D. Martin, A. Lemaitre, J. Bloch, D. N. Krizhanovskii, M. S. Skolnick, C. Tejedor and L. Vina, Nature 457, 291 (2009)

\bibitem{ShelykhJosephson} I.A. Shelykh, D. D. Solnyshkov, G. Pavlovic, and G. Malpuech, Phys. Rev. B \textbf{78}, 041302 (2008)

\bibitem{LagoudakisJosephson} K. G. Lagoudakis, B. Pietka, M. Wouters, R. Andre and B. Deveaud- Pledran, Phys. Rev.
Lett. 105, 120403 (2010)

\bibitem{LaussySupercond} F.P. Laussy, A.V. Kavokin, I.A. Shelykh, Phys. Rev. Lett. \textbf{104}, 106402 (2010)

\bibitem{Imamoglu} A. Imamoglu and J. R. Ram, Phys. Lett. A \textbf{214}, 193, (1996)

\bibitem{Confinement}A. T. Hammack, M. Griswold, L. V. Butov, L. E. Smallwood, A. L. Ivanov, and A. C. Gossard, Phys. Rev. Lett. \textbf{96}, 227402
(2006); R. B. Balili, D. W. Snoke, L. Pfeiffer, and K. West, Appl. Phys. Lett. \textbf{88}, 031110. (2006); O. El Daif, A. Baas, T. Guillet, J.-P. Brantut, R. Idrissi Kaitouni, J. L. Staehli1, F. Morier-Genoud, and B. Deveaud, Appl. Phys. Lett. \textbf{88}, 061105
(2006); R. I. Kaitouni, O. El Daif, A. Baas, M. Richard, T. Paraiso, P. Lugan, T. Guillet, F. Morier-Genoud, J. D. Ganiere, J. L. Staehli, V. Savona, and B. Deveaud, Phys. Rev. B \textbf{74}, 155311 (2006); M. M. Kaliteevskii, S. Brand, R. Abram, I. Iorsh, A. Kavokin, and I. Shelykh,  Appl. Phys. Lett. \textbf{95}, 251108 (2009)

\bibitem{WertzNature} E. Wertz, L. Ferrier, D. Solnyshkov, R. Johne, D. Sanvitto, A. Lemaitre, I. Sagnes, R. Grousson, A. V. Kavokin, P. Senellart, G. Malpuech, and J. Bloch, Nature Physics \textbf{6}, 860 (2010)

\bibitem{LiewNeuron} T.C.H. Liew, A.V. Kavokin and I.A. Shelykh, Phys. Rev. Lett.  \textbf{101}, 016402 (2008)

\bibitem{LiewCircuit} T. C. H. Liew, A. V. Kavokin, T. Ostatnicky, M. Kaliteevski, I. A. Shelykh, and R. A. Abram  Phys. Rev. B \textbf{82}, 033302 (2010)

\bibitem{SpinHall} A.V. Kavokin, G. Malpuech and M. Glazov, Phys. Rev. Lett. \textbf{95}, 136601 (2005)

\bibitem{Glazov} M.M. Glazov and L.E. Golub, Phys. Rev. B \textbf{77}, 165341 (2008)

\bibitem{ShelykhBerry} I.A. Shelykh, G. Pavlovic, D. D. Solnyshkov, and G. Malpuech, Phys. Rev. Lett. \textbf{102}, 046407 (2009)

\bibitem{Bottleneck} F. Tassone, C. Piermarocchi, V. Savona, A. Quattropani and P. Schwendimann, Phys. Rev. B \textbf{56}, 7554 (1997)

\bibitem{polpol} F. Tassone and Y. Yamamoto, Phys. Rev. B \textbf{59}, 10830 (1999)

\bibitem{Carusotto} I. Carusotto and C. Ciuti, Phys. Rev. Lett. \textbf{93}, 166401 (2004).

\bibitem{ShelykhGP} I.A. Shelykh, Yuri G. Rubo, G. Malpuech, D. D. Solnyshkov, and A. Kavokin, Phys. Rev. Lett. \textbf{97}, 066402 (2006)

\bibitem{Porras2002} D. Porras, C. Ciuti, J. J. Baumberg, and C. Tejedor, Phys. Rev. B \textbf{66}, 085304 (2002)

\bibitem{Kasprzak2008} J. Kasprzak, D. D. Solnyshkov, R. Andre, Le Si Dang, and G. Malpuech, Phys. Rev. Lett. \textbf{101}, 146404 (2008)

\bibitem{Haug2005} T.D. Doan, H.T. Cao, D.B.Tran Thoai and H. Haug, Phys. Rev. B \textbf{72}, 085301 (2005)

\bibitem{Cao} H.T. Cao, T. D. Doan, D. B. Tran Thoai, and H. Haug, Phys. Rev. B \textbf{77}, 075320 (2008)

\bibitem{Wouters2007} M. Wouters and I. Carusotto, Phys. Rev. Lett. \textbf{99}, 140402 (2007).

\bibitem{BerloffVortex} M. O. Borgh, J. Keeling, and N. G. Berloff, Phys. Rev. B \textbf{81}, 235302 (2010)

\bibitem{Tim1D} M. Wouters, T. C. H. Liew, and V. Savona, Phys. Rev. B \textbf{82}, 245315 (2010)

\bibitem{Butov} L. V. Butov, \emph{J. Phys.: Condens. Matter} \textbf{19},
295202 (2007).

\bibitem{Josephson2010} E.B. Magnusson, H. Flayac, G. Malpuech and I.A. Shelykh, Phys. Rev. B \textbf{82}, 195312 (2010)

\bibitem{WoutersWigner} M. Wouters and V. Savona, Phys. Rev. B \textbf{79}, 165302 (2009)

\bibitem{Carmichael} H. Carmichael, Quantum Optics 1: Master Equations And Fokker-Planck Equations, Springer, New
York, 2007.

\bibitem{KavokinMalpuech} A. Kavokin and G. Malpuech, Cavity Polaritons (Elsevier
Academic Press, Amsterdam, 2003).


\bibitem{PPO} P.G. Savvidis,  J. J. Baumberg, R. M. Stevenson, M. S. Skolnick, D. M. Whittaker, and J. S. Roberts, Phys. Rev. Lett. \textbf{84}, 1547 (2000)



\end{thebibliography}
\end{document}